\documentstyle[pramana,epsf,floats]{ias}

\def\beq{\begin{equation}}
\def\eeq{\end{equation}}   
\def\bea{\begin{eqnarray}}
\def\eea{\end{eqnarray}}

\begin{document}

\preprint{xxxxxxx, IITB-PH-0002}
\title{Inflation with bulk fields in the Randall-Sundrum warped
compactification?}

\author{J.\ M.\  Cline}
\address{McGill University, 3600 University St.
Montr\'eal, Qu\'ebec H3A 2T8, Canada \\ 
E-mail: jcline@physics.mcgill.ca}

\author{U.\  A.\  Yajnik}
\address{Indian Institute of Technology Bombay, Mumbai 400\thinspace076, 
India \\
E-mail: yajnik@phy.iitb.ernet.in}

\abstract{
Randall and Sundrum have proposed that we live on a brane embedded
in 5-dimensional anti-deSitter space, as a solution to the hierarchy
problem.  We examine the possibility of using a scalar field in the
5-D bulk as the inflaton, and show that it gives results which are
indistinguishable from an inflaton which is restricted to our brane.
This makes it difficult to get large enough density perturbations in
simple one-field inflationary models, since mass scales on our brane
are supposed to be limited by a cutoff of order 1 TeV.}

\maketitle

\section{Introduction}
\label{sec:intro}

The Randall-Sundrum proposal for solving the hierarchy problem has
received much attention in the last year [1].  
They considered an extra
compact dimension with coordinate $y$ and line element
\beq
\label{eq:ds2}
	ds^2 = a^2(y) \eta_{\mu\nu} dx^\mu dx^\nu + b^2 dy^2
\eeq
where $a(y) = e^{-kb|y|}$, $y\in[-1,1]$ and the points $y$ and $-y$ are
identified so the extra dimension is an orbifold with fixed points at
$y=0$ and $y=1$.  The four-dimensional geometry is conformal to
Minkowski space.  The scale $k$ in the warp factor $a(y)$ is determined
by the 5-D cosmological constant, $\Lambda$ and the analog to the
Planck mass, $M$, by $k = (-\Lambda/6M^3)^{1/2}$.  Hence $\Lambda$ must
be negative, and the 5-D space is anti-deSitter.

At $y=0$ there is a positive tension brane (the Planck brane) on which
particle masses are naturally of order the Planck mass, $M_p$, while at
$y=1$ there is a negative tension brane (called the TeV brane), where
particle masses are suppressed by the warp factor $e^{-kb}$.  By
adjusting the size of the extra dimension, $b$, so that $kb\sim 37$, the
masses of particles on the TeV brane will be in the TeV range, even if
all the underlying mass parameters (including $\Lambda$, $M$ and $k$)
are of order $M_p$ to the appropriate power.

Although this is desirable for solving the hierarchy problem, it makes
it difficult to understand the origin of inflation.  The density
perturbations from inflation, $\delta\rho/\rho$, are suppressed by
inverse powers of $M_p$.  To get $\delta\rho/\rho\sim 10^{-5}$, one
needs a mass scale much larger than 1 TeV in the numerator.  By
construction, the TeV scale is the cutoff on the TeV brane, so it is hard
to see where such a scale could come from unless some physics outside of
the TeV brane is invoked.

\section{Inflation with a bulk scalar}

In this study we investigate what happens when the inflaton is a bulk
scalar field.  The simplest possibility is chaotic inflation with a
free field [2].  We will assume that the line element
(\ref{eq:ds2}) is modified by replacing the Minkowski metric with 4-D
deSitter space, whose line element is $ds^2 = -dt^2 + e^{2Ht} d\vec
x^2$.  The action for the bulk scalar is then
\beq
	S = \frac12 \int d^{\,4}x\, dy\, b\, e^{3Ht} a^4(y) \left[ a^{-2}(y)
\dot\phi^2
	- b^{-2} \phi'^2 - m^2\phi^2\right]
\eeq
The equation of motion for $\phi$ is
\beq
	a^{-2}\left(\ddot\phi + 3H\dot\phi\right) - b^{-2}
	\left(\phi''-4kb\phi'\right)+m^2\phi = 0
\eeq
and, assuming that the size of the extra dimension is stabilized
[3], the Hubble rate is given by
\beq
	H^2 = {4\pi G\over 3} \int_0^1 dy\,b\,a^4(y)\left(
	a^{-2}\dot\phi^2 + b^{-2}\phi'^2 + m^2\phi^2\right),
\eeq
where $G$ is the ordinary 4-D Newton's constant.

We look for a separable solution, $\phi = \phi_0(t) f(y)$.  We take
$\phi_0$ to have dimensions of (mass)$^1$ so that it is the canonically
normalized field in an effective 4-D description, and $f$ has dimensions of
(mass)$^{1/2}$.  We will also assume the slow roll condition is fulfilled
so that the terms $\ddot\phi$ and $\dot\phi^2$ can be ignored in the last
two equations.  The equation of motion becomes
\beq
	\dot\phi_0 = {e^{-2kby}\over 3\hat H}\left(
	-m^2 + {1\over b^2 f}\left(f''-4kbf'\right)\right) = 
	{\rm constant} \equiv -\Omega
\eeq
with
\beq
\label{eq:HH}
	\hat H^2 \equiv {H^2\over\phi_0^2} = 
	{4\pi G\over 3} \int_0^1 dy\,b\,e^{-4kby}\left(
         b^{-2}f'^2 + m^2 f^2\right).
\eeq
The solution for $\phi_0$ is obviously linear, $\phi_0(t) = C - \Omega
t$.  For chaotic inflation we want $C \gg M_p$ (necessary to fulfill
the slow roll condition [2]) and $\Omega>0$, so that $\phi_0$
is rolling to the minimum of its potential.

The equation for $f$ becomes
\beq
\label{eq:eom}
	f'' - 4kb f' - b^2\left(m^2 - 3\hat H\Omega e^{2kby}
	\right)f = 0,
\eeq
This is the same equation as (7) of ref.\ [4].  The solutions
$f_n$ (called $y_n$ in [4]) are discrete, such that $\hat H\Omega$
is quantized:
\beq
\label{eq:ev}
	3 k^{-2}\hat H\Omega e^{2kb} = x_{n\nu}^2
\eeq
Here $x_{n\nu}$ is the $n$th root of the equation $2J_\nu(x_{n\nu})
+ x_{n\nu}J'_\nu(x_{n\nu}) = 0$, where the order of the Bessel function
is $\nu=\sqrt{4+m^2/k^2}$.  For $m/k$ in the range $0.5-3$, the lowest
mode $x_{1,\nu}$ ranges from 4 to 6.  This assumes the boundary condition that
$f'=0$ at $y=0,1$, but other choices of boundary conditions will lead to
essentially identical conclusions, as we will explain below.  The $f_n$'s
are normalized so that 
\beq
\label{eq:norm}
	\int_0^1 dy\,b\,e^{-2kby} f_n(y) f_m(y) = \delta_{mn}
\eeq

With the solution for $f_n$ we can evaluate the rescaled Hubble rate,
$\hat H$, in eq.\ (\ref{eq:HH}).  After a partial integration and use
of the equation of motion (\ref{eq:eom}), the integral in (\ref{eq:HH})
becomes identical to that of (\ref{eq:norm}), times $3\hat H\Omega$.
This gives $\hat H^2 = 4\pi G \hat H \Omega$, which together with
eq.\ (\ref{eq:ev}) determines $\Omega$, the rate at which $\phi_0$
is rolling to its minimum:
\beq
	\Omega = {k x_{n\nu} e^{-kb}\over \sqrt{12\pi G}}
	\sim M_p \times {\rm 1 TeV}
\eeq
We used the fact that $e^{-kb}$ is supposed to be of order (TeV)$/M_p$.

\section{Density perturbations}

We can now estimate the magnitude of density perturbations in this model.
Using $\delta\rho/\rho\sim H^2/|\dot\phi_0|$, 
\beq
	{\delta\rho\over\rho}\sim (4\pi G)^2\, \Omega\, (C-\Omega t)^2
	\sim {{\rm TeV}\over M_p}
\eeq
Although $C$ is presumed to be super-Planckian, $(C-\Omega t)$ will not be
orders of magnitude larger than $M_p$ near the end of inflation, when the
perturbations with COBE-scale wavelengths were being produced; hence we
take $(C-\Omega t)\sim M_p$ in the above estimate.   This suppression 
of the density perturbations makes our model not viable.

Nevertheless, it is interesting to compare to what would happen if we
tried to do chaotic inflation using a scalar field trapped on the TeV
brane.  The mass of the field is now constrained to be of order $m
\sim$ TeV because of the suppression of masses by the warp factor.
The equation of motion during the slow roll regime is
\beq
	\dot\phi = -{m^2\phi\over 3H} = - {m^2\phi\over \sqrt{
	4\pi G m^2\phi^2/3} };
\eeq
hence $\phi$ evolves linearly with time, and $\dot\phi$ is of order 
$M_p\times 1 $TeV, just as with the bulk scalar field.  And the estimate
for $\delta\rho/\rho$ has the same parametric form.  All this, despite the
fact that we started with a bulk scalar whose mass is Planck-scale in the
5-D Lagrangian.

In retrospect, this result is not surprising.  Reference [4]
noted that the modes of the bulk scalar behave similarly to TeV-scale
particles on the brane.  This can be understood by the form of the
solutions [4],
\beq
	f_n \sim e^{2ky} J_\nu\left(x_{n\nu}
	e^{kb(y-1)}\right),
\eeq
which are strongly peaked near $y=1$.  There is thus little practical
difference between the low-lying modes of the bulk field and a field
confined to the TeV brane.  

One might wonder if this conclusion depends on the choice of boundary
conditions for the modes $f_n$.  However the fact that the modes peak
at the TeV brane comes from bulk energetics, not boundary conditions.
Since the mass of the bulk field is effectively varying like $e^{-kby}$,
it is energetically much more efficient for the field to be concentrated
near $y=1$.

\section{Conclusion}

The simplest chaotic inflation models seem to be ruled out in the
Randall-Sundrum scenario, whether the inflaton is a bulk field or one
restricted to the TeV brane.  One could, alternatively, put the
inflaton on the Planck brane (at $y=0$), at the cost of reintroducing a
hierarchy problem---why should $m/M_P$ be $O(\delta\rho/\rho)\sim
10^{-5}$?  This fine-tuning problem always occurs in inflation, but the
RS setting casts it in a somewhat new light.  One could invent an
intermediate brane for the inflaton, which has just the right mass
scale, but this seems artificial.  Perhaps the RS idea, if correct, is
telling us that hybrid inflation (involving more than one field) is
necessary.

\end{document}